\documentclass[preprint,superscriptaddress,showkeys,showpacs,floatfix,eqsecnum,onecolumn]{revtex4}

\usepackage{graphics}
\usepackage{subfigure}
\usepackage{palatino}
\usepackage{dcolumn}
\usepackage{bm}
\usepackage{amssymb,amsmath,graphicx,graphics,color,epstopdf}
\usepackage{hyperref}
\usepackage{float} 
\usepackage{changes}
\hypersetup{
    colorlinks=true,
    linkcolor=blue,
    filecolor=magenta, citecolor=blue,    urlcolor=blue,
}

\begin{document}

\title{Quasinormal modes of \textsl{dS} and \textsl{AdS} Black Holes: Feedforward neural network method}
\author{Ali \"{O}vg\"{u}n}
\email{ali.ovgun@emu.edu.tr }
\homepage{https://www.aovgun.com}

\affiliation{Physics Department, Eastern Mediterranean
University, Famagusta, 99628 North Cyprus, via Mersin 10, Turkey.}

\author{\.{I}zzet Sakall{\i}}
\email{izzet.sakalli@emu.edu.tr}
\affiliation{Physics Department, Eastern Mediterranean
University, Famagusta, 99628 North Cyprus, via Mersin 10, Turkey.}

\author{Halil Mutuk}
\email{hmutuk@omu.edu.tr}
\affiliation{Physics Department, Faculty of Arts and Sciences, Ondokuz Mayis University,
55139, Samsun, Turkey.}


\begin{abstract}
In this paper, we show how the quasinormal modes (QNMs)
arise from the perturbations of massive scalar fields propagating in the curved
background by using the artificial neural networks. To this end, we architect a special algorithm for the feedforward neural network method (FNNM) to compute the QNMs complying with the certain types of boundary conditions. To test the reliability of the method, we consider two black hole spacetimes whose QNMs are well-known: $4D$ pure de Sitter (\textsl{dS}) and five-dimensional Schwarzschild anti-de Sitter (\textsl{AdS}) black holes. Using the FNNM, the QNMs of are computed numerically.
It is shown that the obtained QNMs via the FNNM are in good agreement with their
former QNM results resulting from the other methods. Therefore, our method of finding the QNMs can be used for other curved spacetimes that obey the same boundary conditions.

\end{abstract}



\date{\today}

\maketitle

\section{Introduction} \label{intro}

QNMs are single frequency modes dominating the time evolution of perturbations of systems which are subject to damping, either
by internal dissipation or by radiating away energy. Due to the damping, the frequency of a QNM must be complex,
its imaginary part being inversely proportional to the typical damping time. Recall that, in general relativity, damping occurs even without
friction, since energy may be radiated away towards infinity by gravitational waves \cite{IP1,IP2}. Thus, they could lead to the \textit{direct} identification of the existence of
the BH through gravitational wave observation, which might be realized
in the near future. QNMs, which are believed to be
characteristic sounds of perturbed black hole (BH) spacetimes have been investigated for a long time and their physical properties have been presented in various studies (see for example
\cite{Schutz:1985zz,Berti:2009kk,Guinn:1989bn,Konoplya:2011qq,Iyer:1986nq,CastelloBranco:2005hz,CastelloBranco:2004nk,Jing:2005dt,Skakala:2010uk,Sakalli:2013yha,Sakalli:2011zz,Sakalli:2018nug,Ovgun:2018gwt,Gonzalez:2017zdz,Jusufi:2017trn,Ovgun:2017dvs,Crisostomo:2004hj,Lepe:2004kv,Saavedra:2005ug,Becar:2007hu,Becar:2010zz,Abdalla,Casals:2012tb,Konoplya:2005hr,Kuang:2017sqa,Jusufi:2015mii}
). It is worth noting that similar to the BH geometries, QNMs can also arise from other spacetimes including the wormholes
\cite{Konoplya:2005et,Kim:2008zzj,Konoplya:2016hmd,Oliveira:2018oha,Sakalli:2015taa,Sakalli:2015mka,Ovgun:2015sqa,Ovgun:2016ijz,Jusufi:2017vta,Halilsoy:2013iza,Jusufi:2017mav,Richarte:2017iit,Ovgun:2018fnk,Konoplya:2018ala,Kim:2018ang,Bronnikov:2012ch,Aneesh:2018hlp,Konoplya:2010kv,Bueno:2017hyj,Volkel:2018hwb,Aragon:2020xtm,Yan:2020hga,Fontana:2020syy,Panotopoulos:2020mii,Mourier:2020mwa,Fabris:2020kog,Villani:2020bfc,Chen:2020evr,Burikham:2020dfi,Matyjasek:2020bzc,Aragon:2020tvq,Awad:2017sau,Hanafy:2015yya,Nashed:2004pn}. 

\bigskip

Since the discovery of the \textsl{AdS/CFT} correspondence \cite{IZS1}, QNMs of \textsl{AdS} spacetimes has become very attractive over the past two decades.
Besides, it is suggested that QNMs of \textsl{AdS} BHs are related with the double conformal field theory (CFT)
\cite{IZS4,IZS5,IZS6,IZS7,IZS8,IZS9}. Since the QNMs govern the deterioration time
of a perturbed BH, within the bulk, configuration, they should be
associated with the \textsl{AdS/CFT} duality in order to return the boundary of
Yang-Mills theory to the thermal equilibrium. The numerical computations of QNMs for \textsl{AdS} BHs in arbitrary dimensions were served in
\cite{IZS4}. Then, Govindarajan and Suneeta \cite{IZS10} computed the QNMs of
the $5D$ \textsl{AdS}-Schwarzschild BH by using the superpotential
approach. Moreover, in the framework of scalar perturbation spectra, it was
known that there exists a relation between (bulk) \textsl{dS} spacetime and the
corresponding CFT at the boundaries (\textit{past} $\textsl{I}^{-}$ and \textit{future} $\textsl{I}^{+}$)
\cite{IZS11}, which provides a quantitative support for the dS/CFT
correspondence. The relation between the QNMs and
surface gravity ($\kappa$) of the cosmological horizon was thoroughly discussed in \cite{IZS12}.
Unlike the massless minimally coupled scalar field, it was shown that for a
massive scalar field there exists QNMs in the pure \textsl{dS} spacetimes. Even, the
obtained QNMs of pure \textsl{dS} spaces are analytical frequencies \cite{IZSMAIN}.

New derivations of the QNMs for the curved spacetimes have always attracted a
great attention. This challenge stems from the fact that it is difficult to
solve the wave equations of the considered fields, analytically. Therefore,
many numerical techniques have been developed in order to solve those type of
equations. In recent decades, the artificial neural networks (ANNs) \cite{ann2019}
are employed for finding solutions of differential equations which appeared in the
different physical systems. As is well known, FNNM was the first and simplest type of ANN devised. In this network, the information moves in only one direction forward from the input nodes, through the hidden nodes (if any) and to the output nodes. There are no cycles or loops in the network   \cite{FNNM}. FNNM or such connectionist systems
compute the systems uncertainly inspired by the biological nervous (neural) systems that
constitute animal brains and also there are many applications in general relativity and cosmology \cite{Menendez-Vazquez:2020khz,Dreissigacker:2020xfr,Wang:2020dbt,Vajente:2019ycy,Ciuca:2018tei,Khan:2020fso,Green:2020dnx,Santos:2020gis,Haegel:2019uop,Dreissigacker:2019edy,George:2017pmj,Gabbard:2017lja}. Some of the advantages of using FNNM in solving the
differential equations are listed below \citep{Parisi:2003,Yadav:2015,Mutuk:2018erw,Mutuk:2019uez}:

\begin{itemize}
\item Solution in the domain/field of integration is continuous
\cite{Nakano:2018vay},

\item Computing complexity does not increase significantly with increasing number
of sampling points and dimensions,

\item Rounding-off error propagation does not alter the ANN solution, which happens in standard numerical methods

\end{itemize}

In this paper, we separately compute the QNMs for the four-dimensional pure \textsl{dS}
space \cite{dSiz1,IZSMAIN} and the $5D$ \textsl{AdS}-Schwarzschild BH \cite{Govindarajan:2000vq} by using the FNNM within the framework of \textit{supervised learning} (see for instance \cite{SLearning} and references therein). In fact, the supervised learning algorithm analyzes some training (educational) data and generates an inference (or the so-called \textit{trial}) function that can be used to achieve new results. 
This requires that the learning algorithm should be reasonably generalized from the educational data to situations that are not visible. It was also shown in \cite{accuracy} that  the accuracy of the results obtained from the neural network surpasses the accuracy of other machine learning algorithms like SVM (\textsl{support vector machines}) and RF (\textsl{random forest}). While performing the
computations, we consider the massive scalar field perturbations of the
associated spacetimes. Finally, we compare the QNM values obtained with the FNNM with
the results found from the other methods.

This paper is organized as follows: In \hyperref[II]{Sec. II}, we briefly review the
QNMs of the $4D$ pure \textsl{dS} and the $5D$
\textsl{AdS}-Schwarzschild BH. In \hyperref[III]{Sec. III}, we describe the FNNM and show how one can compute the QNMs of those \textsl{dS/AdS} spacetimes. Then, we present and compare our results with the
known ones. Finally, we conclude the paper with discussions in \hyperref[conc]{Sec. IV}. We use natural units with $G=\hbar=c=1$.

\section{QNMs of Pure \textsl{dS} and \textsl{AdS}-Schwarzschild BHs} \label{II}

In this section, we shall make a brief overview of the QNMs of the pure \textsl{dS} and
\textsl{AdS} spacetimes, which were obtained by the methods of analytical and
super potential approach, respectively. We first consider the pure \textsl{dS}
spacetime, which is given by the following $4D$ line-element
\cite{dSiz1}:
\begin{equation}
ds^{2}=-f(r)dt^{2}+f^{-1}(r)dr^{2}+r^{2}d\Omega_{2}^{2},\label{new1}%
\end{equation}
where $f(r)=1-(\frac{r}{l})^{2}$ in which $l$ denotes the minimal
radius of \textsl{dS} space. Furthermore, $r^{2}d\Omega_{2}^{2}$ is the metric
on the $2D$ sphere of radius $r$.
For the massive scalar field $\Phi$ perturbations, one should consider the
Klein-Gordon equation:
\begin{equation}
{\Phi^{;\nu}}_{;\nu}=m^{2}\Phi,\label{new2}%
\end{equation}
which can be separated by
\begin{equation}
\Phi=\frac{\Psi^{dS}(r)}{r}e^{-i\omega t}Y_{\ell}%
(\Omega_{2}). \label{newiz2} 
\end{equation}
Here, $Y_{\ell}(\Omega_{2})$ is nothing but the spherical
harmonics, which corresponds to the eigenfunction of two-dimensional
Laplace-Beltrami operator $\nabla_{2}^{2}$ having the eigenvalue $-\ell
(\ell+1)$. Recalling definition of the tortoise coordinate, we get%

\begin{equation}
r_{\ast}=\int\frac{dr}{f(r)}=l\tanh^{-1}\bigg(\frac{r}{l}\bigg).\label{new3}%
\end{equation}

Thus, one obtains the radial equation in the form of $1D$
Schr\"{o}dinger-like wave equation %

\begin{equation}
-\frac{d^{2}\Psi^{dS}}{{dr_{\ast}}^{2}}+\left[  V^{dS}_{0}(r)-\omega^{2}\right]
\Psi^{dS}=0,\label{new4}%
\end{equation}

where the effective potential reads
\begin{equation}
V^{\textsl{dS}}_{0}(r)=\frac{1}{l^{2}}\left[  \frac{\ell(\ell+1)}{\sinh^{2}(r_{\ast}/l)}%
-\frac{2-m^{2}l^{2}}{\cosh^{2}(r_{\ast}/l)}\right]  ,\label{new5}%
\end{equation}

Since $\ell(\ell+1)\geq0$, the effective potential (\ref{new5}) diverges ($\rightarrow\infty$) at the singularity ($r=0$) and vanishes at the cosmological
horizon ($r_{h}$). For this reason, QNMs obey the following boundary
conditions: purely outgoing wave at the cosmological horizon and vanish at the
singularity \cite{IZSMAIN}. Meanwhile, at this stage, it is worth noting the late-time tails \cite{Burko:2007ju} cannot be addressed by merely studying Eq. \eqref{new4} (the reader can refer to \cite{Destounis:2020pjk}). After deriving the exact analytical solution of
the radial equation in terms of the hypergeometric function and in the sequel
imposing the boundary conditions, it was found that to have non-zero QNMs
there is a lowest bound: $m>\frac{3}{2l}$ on the mass of scalar field ${\Phi}$. The
resulting QNM sets were given by \cite{IZSMAIN} as follows:
\begin{align}
\omega_{I}  & =\pm\frac{1}{l}\big[m^{2}l^{2}-\frac{9}{4}\big]^{\frac{1}{2}%
}-\frac{i}{l}(2n+\ell+\frac{3}{2})\label{new6}\\
\text{or}\qquad\,\omega_{II}  & =\pm\frac{1}{l}\big[m^{2}l^{2}-\frac{9}%
{4}\big]^{\frac{1}{2}}-\frac{i}{l}(2n-\ell+\frac{1}{2})\label{new7}%
\end{align}

The difference in sets is due to the poles of the gamma functions that help us
to sort out the waves on the horizon only as outgoing waves. Without loss of
generality, when comparing the above results with the FNN method to be
applied, we will consider the first set as%

\begin{equation}
\omega=\frac{1}{l}\big(m^{2}l^{2}-\frac{9}{4}\big)^{\frac{1}{2}}-\frac{i}%
{l}(2n+\ell+\frac{3}{2}).\label{new8}%
\end{equation}

On the other hand, $5D$ \textsl{AdS}-Schwarzschild BH is given by
\cite{Govindarajan:2000vq}
\begin{equation}
ds^{2}=-N(r)dt^{2}+N^{-1}(r)dr^{2}+r^{2}d\Omega_{3}^{2},\label{new9}%
\end{equation}

where
\begin{equation}
N(r)=1+\bigg(\frac{r}{l}\bigg)^{2}-\bigg(\frac{r_{0}}{r}\bigg)^{2}.\label{new10}%
\end{equation}
The relationship between $r_{0}$ and the BH mass $M$ is given by
\begin{equation}
M=\frac{3A_{3}r_{0}^{2}}{16\pi G_{5}},\label{new11}%
\end{equation}
where $A_{3}$ denotes the area of a unit $3D$-sphere described by $d\Omega
_{3}^{2}$. Using the ansatz for the scalar field
\begin{equation}
\Phi=r^{-3/2}\Psi^{\textsl{AdS}}(r)\exp(i\omega t),\label{new12}%
\end{equation}
the massless Klein-Gordon equation of metric (\ref{new9}) yields the following one-dimensional
Schr\"{o}dinger-like wave equation:%

\begin{equation}
-\frac{d^{2}\Psi^{\textsl{AdS}}}{{dr_{\ast}}^{2}}+\left[  V^{\textsl{AdS}}_{0}(r)-\omega^{2}\right]
\Psi^{\textsl{AdS}}=0,\label{new13}%
\end{equation}

where $dr_{\ast}=\frac{dr}{N(r)}$ and
the effective potential becomes (for simplicity, the authors of
\cite{Govindarajan:2000vq}\ had taken $l=1$ and we will also stick to their
choice in our computations in order to make a consistent comparison):
\begin{equation}
V^{\textsl{AdS}}_{0}(r)=N(r)\left(  \frac{15}{4}+\frac{3}{4r^{2}}+\frac{9r_{0}^{2}}{4r^{4}%
}\right)  .\label{new14}%
\end{equation}

It is clear from Eq. (\ref{new14}) that $V^{\textsl{AdS}}_{0}\rightarrow\infty$ at spatial infinity ($r=\infty$) and vanishes at the horizon
$\left[  r=r_{+}\text{ }\rightarrow\text{\ }N(r_{+})=0\right]  $. For this
reason, QNMs of obey the following boundary condition: purely ingoing wave at
the horizon and vanish at the spatial infinity. Since the fundamental QNMs of
the Schwarzschild BH are closely approximated by the QNMs of the
P\"{o}schl-Teller potential, in the spirit of the P\"{o}schl-Teller method for
asymptotically flat BHs, the QNMs for the \textsl{AdS}-Schwarzschild BH
in $5D$, using a superpotential approach, was obtained and served
in Table I of \cite{Govindarajan:2000vq}. In general, for the asymptotically flat BHs, the QNMs correspond to solutions of the wave equations with the physical boundary conditions of purely outgoing waves at spatial infinity and purely ingoing waves crossing the event horizon \cite{myQNM1,myQNM2}. However, for the \textsl{AdS}-Schwarzschild BH, QNMs should admit wave functions that must be purely ingoing wave at the horizon and no outgoing wave at spatial infinity. Namely, at the asymptotic regions, all QNMs of the \textsl{AdS} BH
are required to terminate. Thus, as being
highlighted in \cite{Govindarajan:2000vq}, any numerical calculation of QNMs is
very \emph{artful} due to the nature of the boundary conditions.During the numerical computations, one must ensure to have pure ingoing wave near the horizon, which could be contaminated by an outgoing wave and the
correct \emph{asymptotic} behavior of the wave function that fades away as
$r\rightarrow\infty$. In the superpotential method  \cite{Govindarajan:2000vq}, as in the continued
fraction method \cite{CFMiz} which is suitable for the asymptotically flat
BHs, a particular ansatz for the wave function was introduced to meet
all boundary conditions. Similarly, in the FNNM method, a trial solution
or ansatz that meets the boundary requirements will have to be sought.

\section{FNNM} \label{III}
A complicated problem in science can be solved analytically or numerically in terms of known methods. In most of the cases, an analytical solution to the associated differential equation may not be obtained easily and it is usually cumbersome. Various types of numerical methods have been developed to solve such transcendental differential equations such as shooting, Euler, Runge-Kutta, finite difference, finite element, finite volume, Adomian decomposition, asymptotic iteration, variational iteration, and perturbation methods. All these methods have both advantages and shortcomings. Although they provide good approximate solutions, these methods require discretization of the domain of the problem. Most of these numerical methods give solutions over discrete points and the solution between these points needs to be interpolated. Besides, these methods are in general iterative such that one should fix the step-size before solving the considered problem. The advantages of employing the ANNs (and whence the FNNM) can be listed as follows \citep{Yadav:2015, Parisi:2003}
:
\begin{itemize} 
    \item Solving differential equation by neural network framework presents solution with a very good generalization properties.
    \item The method is general and can be applied to the systems defined on either orthogonal box boundaries or on irregular arbitrary shaped boundaries.
    \item The ANN method can be implemented on parallel architectures which can be used in more complex problems.
    \item The ANN method spends negligible computing time and memory.
    \item If the model has free parameters, they can be treated as variables in the ANN method.
\end{itemize}

Most of the problems which can not be solved analytically are turned into an mathematical optimization problem in which a numerical solution is sought. This optimization can be done in some techniques. Since the problem is considered in a specific region (i.e., the convergence problem), the transition between local and global solutions require intensive processing. In this perspective, the ANNs have broad usage field and they are powerful tools for performing a mathematical modeling.

An ANN can be defined as  parallel information processor in which a number of neurons are distributed as operating units. This information processing systems can take many input from outside, combines them via mostly  nonlinear operations and produce the output. Nowadays, ANN is one of the popular topics of machine learning paradigm. They have a wide range of usage from pattern recognition to financial forecast
including classification, decoding speech etc. ANNs are typically composed of layers. These layers are made of interconnected neurons (perceptrons in modern computers). A neuron is the main processing element in the ANN. Because, neurons have activation functions which translate input signals to
output signals. Problem solving process in the ANN occurs by acquiring
knowledge. This mechanism is maintained by learning methods and information is
stored within 'inter-neuron connections' strength which can be calculated by
some numerical values called weights \cite{Chak,mtk}.

\begin{figure}[h]
\centering
\includegraphics[width=15cm]{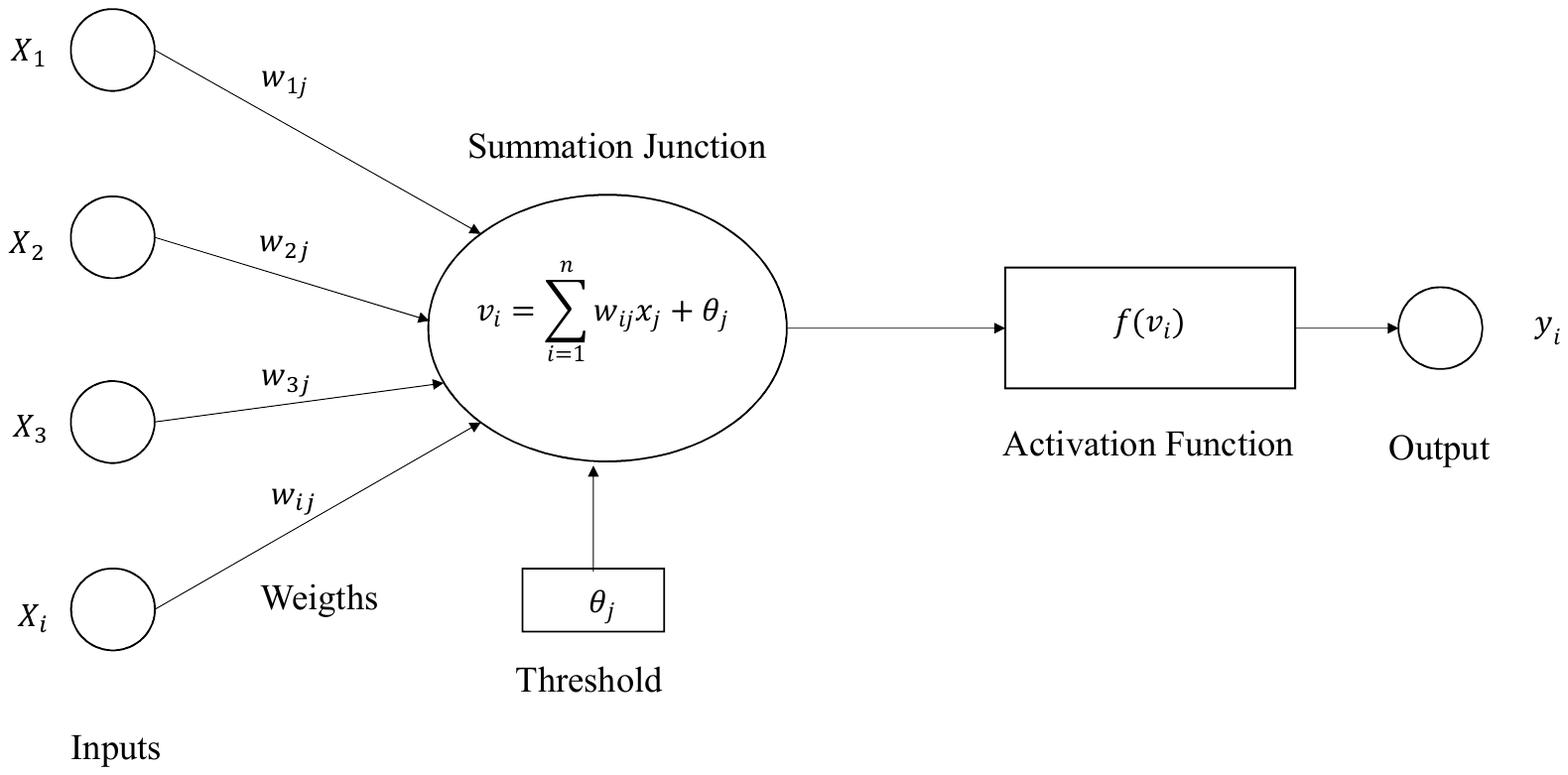} \caption{Basic model of multi-inputs one-output neuron}%
\label{fig:single}%
\end{figure}

 The detailed computational steps of the working principle of an
artificial neuron in a neural network can be seen in in Fig. \eqref{fig:single}.

A neuron $N_{i}$ receives inputs  $n$ which belong to $S=\left\lbrace x_{j} \vert
j=1,2,\cdots,n \right\rbrace $. Each input is multiplied by a weight factor $w_{ij}$ for $j=1,2,\cdots,n$  before  entering to the neuron $N_{i}$. In general, this neuron has a bias term $w_{0}$ and a critical value $\theta_{k}$. To produce the output signal, this critical value must be reached and / or exceeded. The input of the  $i$-th neuron $N_{i}$ in the input layer can be written as
\begin{equation}
v_{i}=w_{0}+\sum_{k=1}^{n} w_{ik}x_{k}.
\end{equation}
The neurons in the input layer can work only if the signal reaches/exceeds the critical value which can be defined as the neuron's working condition as
\begin{equation}
w_{0}+ \sum_{k=1}^{n} w_{ik}x_{k} \geq\theta.
\end{equation}
All the input signals are multiplied by their synaptic weights and added together. This compose "`net"' input to the neuron:
\begin{equation}
\text{net}=\sum_{k=1} w_{ik}x_{k}+\theta,
\end{equation}
where $\theta$ is the threshold (i.e., critical) value. The output signal of $i$-th neuron $N_{i}$ the  can be functionalized as
\begin{equation}
O_{i}= (w_{0}+ \sum_{k=1}^{n} w_{ik}x_{k}).
\end{equation}
An activation function acts on the produced weighted signal which is denoted as $\sigma(s)$. The output signal $y$ can be obtained by mapping this activation function as 
\begin{equation}
y=\sigma(net)=\sigma\left(  \sum_{j=1}^{n} w_{ij}x_{j}+\theta\right)
\end{equation}
where $\sigma$ denotes the neuron activation function. This output function is suggested together with a critical function. In this present work, we will use a sigmoid activation function 
\begin{equation}
\sigma(x)=\frac{1}{1+e^{-x}} \label{sigmoid}
\end{equation}
which is a traditional one for obtaining solutions in nonlinear problems. 

\begin{figure}[h]
\centering
\includegraphics[width=12cm]{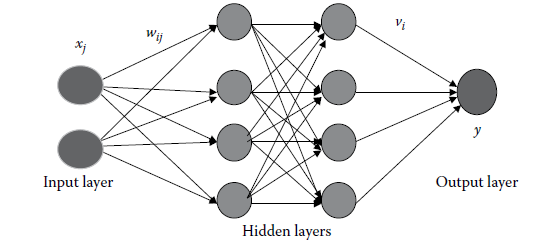} \caption{A sample architecture of an
ANN.}%
\label{fig:ann1}%
\end{figure}

A diagram of a multilayer ANN is given in Fig.
\eqref{fig:ann1}.

The information are given to the input layer, which sends
the information to the hidden layer, if any exists. The processing of inputs
are being done at this stage via a system of weighted connections. The hidden
layers send the information to the output layer and an answer is given to the
outside world. In Fig. \eqref{fig:ann1}, $x_{j}$ are input nodes, $\omega
_{ij}$ are weights from input to the hidden layer (or layers if exist), and $\nu_{i}$ are
synaptic weights from hidden to the output layer $y$ which is the output
node \cite{Chak}. The neurons in the same layer have no connection among themselves.If there is more than one hidden layer, the architecture is known as deep neural network which is out of the scope of the present work.

\begin{figure}[h]
\centering
\includegraphics[width=15cm]{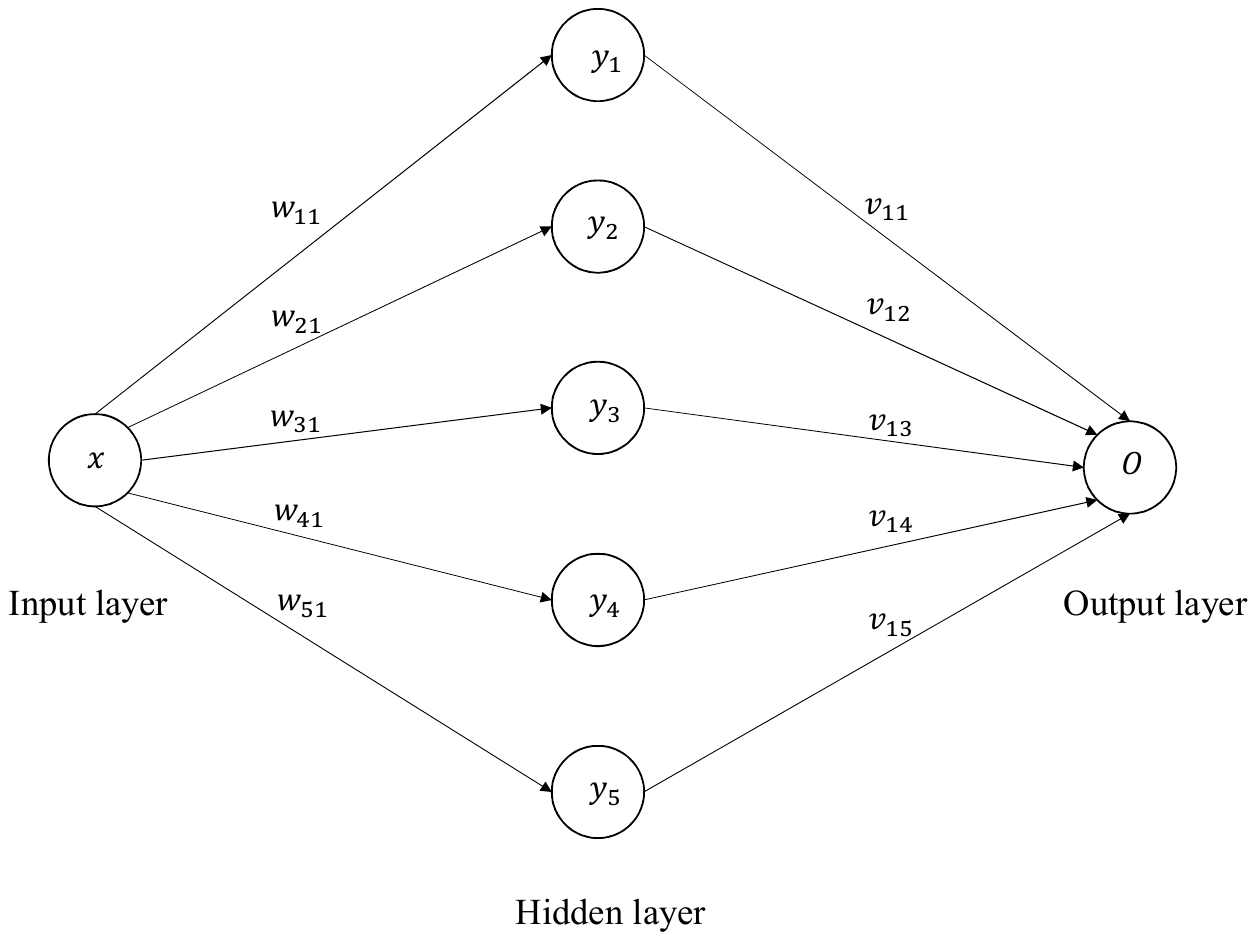} \caption{ANN architecture of this work}%
\label{fig:arch}%
\end{figure}

In the present work we have used an architecture which consists of one input layer, one hidden layer and one output layer. This ANN architecture can be seen in Fig.  \ref{fig:arch}. Neurons are arranged into distinct layers with each layer receiving input from the previous layer and outputting to the next layer. In this manner, neurons (processing elements) in a layer receive input from the previous layer and send (feed) their output to the next layer.

Initial weights from the input layer to the hidden layer ($w_j$) and from the hidden layer to the output layer ($v_j$) are taken as arbitrary
(random). The number of nodes in the hidden layer are determined by the trial-and-error method.

\subsection{Implementation of ANN on Quantum Systems}

The framework for FNNM to obtain solutions of eigenvalue equations was developed in
Ref. \cite{Lagaris:1997ap}. A differential equation can be written as
\begin{equation}
H\Psi(r)=\wp(r), ~ \text{in} ~ \mathcal{D}  \label{eqn2}%
\end{equation}
with
\begin{equation}
\Psi(r)=0 ~  \text{on} ~ \partial \mathcal{D}.
\end{equation}
Here, $H$ is a linear operator, $\wp(r)$ is a \textit{known} function and $\partial \mathcal{D}$ is the boundary of $\mathcal{D}$. In order to solve Eq. \eqref{eqn2}, a trial function 
\begin{equation}
\Psi_{t}(\vec{r})=A(\vec{r})+B(\vec{r}, \vec{\lambda})N(\vec{r}, \vec{p})
\end{equation}
can be used. This function proceeds to a neural network with a vector parameter $\vec{p}$ and undetermined parameter $\vec{\lambda}$ which is going to be adjusted later. $N(\vec{r}, \vec{p}) $ is a single-output feed forward neural network with parameters $\vec{p}$ and $n$ input units fed
with the input vector $\vec{r}$. The functions $A$ and $B$ will be  determined with respect to appropriate $\Psi_{t}(\vec{r})$ which satisfies the boundary conditions.

To solve Eq. \eqref{eqn2}, the \textit{collocation} method \cite{collo} can be used. The idea behind the collocation method is to choose a finite-dimensional space of trial solution functions and a number of points in the domain of the problem, and to select that trial solution which satisfies the given equation at the collocation points. This procedure discrete the domain into a set of $\vec{r}_i$. To this end, one can get a minimization problem as follows:

\begin{equation}
\underset{p,\lambda}{\min}\sum_{i}\left[  H\Psi_{t}(r_{i})-\wp(r_{i})\right]
^{2}.\label{dif2}%
\end{equation}
To have the Schr\"{o}dinger equation, one can recast Eq. \eqref{eqn2} in an
eigenvalue equation:
\begin{equation}
H\Psi(r)=\omega^{2}\Psi(r)
\end{equation}
with the boundary condition $\Psi(r=0)=0$. Before obtaining QNMS via FFNM, an important notice should be done. In this work, we assume the foreknown definition of the QNMs that these modes are purely
outgoing waves at the event horizon of black hole $r_{h}$, where is the boundary defining the region of space around a black hole from which nothing (not even light) can escape and vanish at $r=0$ \cite{IZSMAIN}.
These boundary conditions are determined by the behavior of the effective potential: recall Eqs. \eqref{new5} and \eqref{new14}, which are obtained for \textsl{dS} and \textsl{AdS} spacetimes, respectively. Thus, the trial solution becomes
\begin{equation}
\Psi_{t}(r)=B(\vec{r},\vec{\lambda})N(\vec{r},\vec{p})
\end{equation}
where $B(\vec{r},\vec{\lambda})=0$ at boundaries for a range of
$\lambda$ values. Employing the  discretization for the domain of the problem together with the collocation method, a minimization problem can be obtained with respect to the $\vec{p}$ and
$\vec{\lambda}$:
\begin{equation}
E(\vec{p},\vec{\lambda})=\frac{\sum_{i}\left[  H\Psi_{t}(r_{i}%
,\vec{p},\vec{\lambda})-\omega^{2}\Psi_{t}(r_{i},\vec{p}%
,\vec{\lambda})\right]  ^{2}}{\int|\Psi_{t}|^{2}d\vec{r}},
\end{equation}
where $E$ represents the error function. Furthermore, $\omega^{2}$ is obtained as follows
\begin{equation}
\omega^{2}=\frac{\int\Psi_{t}^{\ast}H\Psi_{t}d\mathbf{r}}{\int|\Psi_{t}%
|^{2}d\mathbf{r}}.
\end{equation}
Thus, the energy eigenvalues or the QNMs are given by
\begin{equation}
\omega=\Bigg(\frac{1}{\int|\Psi_{t}|^{2}d\vec{r}}\left[  \int_{r_{1}%
}^{r_{2}}\Big(\frac{d\Psi_{t}}{dr}\Big)^{2}dr+\int_{r_{1}}^{r_{2}}V_{0}%
(r)\Psi_{t}^{2}(r)dr\right]  \Bigg)^{\frac{1}{2}}.\label{best}%
\end{equation}

where $r_{2}$ represents the location where the scalar field becomes pure
plane wave (ingoing/outgoing) and $r_{1}$ indicates the radial position where
the effective potential diverges and whence causes the waves to be completely
damped (i.e., $\Psi_{t}=0$). Therefore, while $r_{1}=0$ and  $r_{2}=r_{h}$ for
the pure \textsl{dS} BH, in the \textsl{AdS} BH we take $r_{1}=\infty$ and
$r_{2}=r_{+}.$ $V_{0}$ corresponds to the effective potential of the spacetime taken into account.

The parameter $\vec{p}$ is nothing but the weights and biases of the ANN. Although the multi-layer sensor (MLP) has many hidden layers, here we use a simple model of a single hidden layer MLPs.  In this study, we also consider a multilayer perception with $n$ input units, one
hidden layer having $m$ units, and an output. Given an input vector
\begin{equation}
\vec{r}=\left(  r_{1}, \cdots,r_{n} \right),
\end{equation}
the output of the neural network can be written as
\begin{equation}
N=\sum_{i=1}^{m} \nu_{i} \sigma(z_{i}),
\end{equation}
where
\begin{equation}
z_{i}=\sum_{j=1}^{n} \gamma_{ij}r_{j}+u_{i}.
\end{equation}
Here, $\gamma_{ij}$ are the weights from input unit $j$ to the hidden unit
$i$, $\nu_{i}$ is the weight from hidden unit $i$ to the output unit, $u_{i}$
represents the bias of hidden unit $i$, and $\sigma(z)$ is the sigmoid
function, which is given in Eq. \eqref{sigmoid}. The derivatives of the ANN output can be written as
\begin{equation}
\frac{\partial^{k} N}{\partial r^{k}_{j}}=\sum_{i=1}^{m} \nu_{i} \gamma
_{ij}^{k} \sigma_{i}^{(k)},
\end{equation}
where $\sigma_{i}=\sigma(z_{i})$ and $\sigma^{(k)}$ is the $k^{th}$ order
derivative of the activation (sigmoid) function.

One can parametrize the solution trial function as
\begin{equation}
\phi_{t}(r)=e^{-\beta r^{2}}N(r,\vec{u},\vec{w},\vec{v}),~\beta>0,
\end{equation}
where $N$ denotes a feedforward neural network with one hidden layer and $m$
sigmoid hidden units
\begin{equation}
N(r,\vec{u},\vec{w},\vec{v})=\sum_{j=1}^{\Bar{m}}\nu_{j}\sigma
(\omega_{j}r+u_{j}).
\end{equation}
The minimization problem turns out to be as
\begin{equation}
\frac{\sum_{i}\left[  H\phi_{t}(r_{i})-\omega^{2}\phi_{t}(r_{i})\right]  ^{2}%
}{\int|\phi_{t}(r)|^{2}dr}.
\end{equation}

Solving this equation is equivalence to solving Schr\"{o}dinger equation. The minimization problem can be solved via collocation method. In this method, one chooses a finite dimensional space of solution trial function which is supposed to solve given differential equation with a number of points in the domain.

In order to obtain desired result, ANN needs to learn. Solving a differential equation within ANN method requires training of the ANN. This learning process can be done in different ways. In this work, we used error back-propagation learning algorithm. This learning algorithm is one of the most common used learning rules and it is valid for continuous activation function such as sigmoid function Eq. \eqref{sigmoid}.
By taking the partial derivative of the error function according to each weight, we can monitor the flow of the error direction in the network. The steps for the learning algorithm of back-propagation is as follows \cite{Chak,mtk}:

\begin{enumerate}
\item[1] Set the weights $w$ and $v$ from the hidden to the output layer. Choose the
learning parameter in the range $(0,1)$, and error $E_{max}$. At the first step, error is taken to be zero.
\item[2] Train the network.
\item[3] Find the output of error function.
\item[4] Calculate the error signal terms by output and hidden layers, respectively.
\item[5] Calculate the error components for gradient vectors.
\item[6] Check if weights are adjusted appropriately.
\item[7] If $E=E_{max}$, then cease the training. If not, proceed to step 2 by setting $E \to 0$ and initiate the new training.
\end{enumerate}

The crucial point for the training process is taking the eigenvalue (error function) $E$ as zero and train the neural network with equidistant points in the given interval of the problem. It is expected that this process yields energy function (eigenvalue) to be zero or at least converge to zero. If the convergence is not obtained, then the eigenvalue is wrong. If this happens, eigenvalue should be changed in a proper way and the training process should be restarted. It should be keep doing this process until the energy (error) function converges to zero.

It should also be noted that the method for solving differential equations with ANNs does not depend on the training method: The choice of training method only effects the speed of the training procedure.

\subsection{QNMs of Pure \textsl{dS} and \textsl{AdS}-Schwarzschild BHs via FNNM} 

To derive the QNMS of the pure \textsl{dS} and \textsl{AdS}-Schwarzschild BHs, we first consider Eqs. \eqref{new5} and \eqref{new14}, respectively, in Eq. \eqref{best}, then compute the QNMs with the expression seen in Eq. \eqref{best}. To this end, we  employ the Gauss-Legendre rule \cite{Lagaris:1997ap} and use 200
equidistant points in the interval $0<r<10$ with $\Bar{m}=10$. In Tables I and II, we represent our findings, which are the numerical values (via the FNNM within the context of supervised learning) of the QNMs of the pure \textsl{dS} and \textsl{AdS} spacetimes. It can
be seen from those Tables that FNNM satisfactorily re-derives the
well-accepted QNMs' results obtained from the other methods
\cite{IZSMAIN,Govindarajan:2000vq}. Thus, we have managed to
introduce a new and effective method to the literature for computing the QNMs.

\begin{table}[H]
\begin{ruledtabular}
\begin{tabular}{ccccccc}
$(n,\ell)$ & This work &  Ref. \cite{IZSMAIN} & PE (\text{\%})
\\
\hline
$(0,0)$ & $0.-3017i $ & $0.-3000i$ &0.56 \\  
$(0,1)$ & $0.-3994i $ & $0.-4000i$ & 0.15\\
$(1,1)$ & $0.-6012i $ & $0.-6000i$ & 0.2\\
$(0,2)$ & $0.-5024i $ & $0.-5000i$ & 0.48\\
$(1,2)$ & $0.-7019i $ & $0.-7000i$ &0.27 \\
$(2,2)$ & $0.-9037i $ & $0.-9000i$ & 0.41\\
$(0,3)$ & $0.-6072i $ & $0.-6000i$ &1.2 \\
$(1,3)$ & $0.-8056i $ & $0.-8000i$ &0.7\\
$(2,3)$ & $0.-10094i $ & $0.-10000i$ &0.94\\
$(3,3)$ & $0.-12103i $ & $0.-12000i$ &0.85 \\
\end{tabular}\label{Tab1}
\end{ruledtabular}
\caption{Comparison of FNNM QNMs with numerical QNMs obtained, via the superpotential approach method, for the \textsl{AdS} spacetime \cite{Govindarajan:2000vq} (for $l=0.001$ \textit{case}).  Percentual error (PE) rates are given.}%
\end{table}

\begin{table}[H]
\begin{ruledtabular}
\begin{tabular}{ccccccc}
$\text{Radius}, r+$ & This work & Ref. \cite{Govindarajan:2000vq} & PE (\text{\%}) Real& PE (\text{\%}) Imaginary&
\\
\hline
$1$ & $0.6960+1.4629i $  & $0.6948+1.4648i$ & 0.17 &0.12  \\
$2$ & $1.0774+1.9849i $  & $1.0713+1.9817i$ & 0.56 &0.16 \\
$5$ & $2.4407+4.2689i $  & $2.4462+4.2642i$ & 0.22 &0.11 \\
$10$ & $4.8205+8.3285i $  & $4.8249+8.3279i$ & 0.09 &0.07\\
$50$ & $24.0731+41.3784i $  & $24.0159+41.3183i$ & 0.23 &0.04 \\
$100$ & $48.0376+82.6577i $  & $48.0251+82.6165i$ & 0.02 &0.01\\
$150$ & $72.0373+123.9374i $  &  $72.0358+123.9190i$ & 0.06 &0.01\\
$500$ & $240.1166+413.0576i $  & $240.1150+413.0500i$ & 0.01 &0.01\\
$750$ & $360.1145+619.5867i $  &  $360.1720+619.5740$& 0.01 &0.01\\
$1000$ & $480.2143+826.0761 $  & $480.2290+826.0980$& 0.03 &0.02\\
\end{tabular}\label{Tab2}
\end{ruledtabular}
\caption{Comparison of FNNM QNMs with analytical QNMs obtained for the pure \textsl{dS} spacetime \cite{IZSMAIN}. PE rates are also shown.}%
\end{table}

As mentioned above, the main advantage of using ANN is to solve the Schr\"{o}dinger equation. However, the computational complexity while using the ANN does not increse considerably with the number of sampling points and with the number of dimensions in the problem. Depending on the learning algorithm, the running CPU time can be lowered significantly to obtain the solution. On the other hand, we shall not perform any CPU time comparison in this study, because it is irrelevant with the scope of the paper.

\section{Conclusions} \label{conc}

In this study, we have prescribed a new method, FNNM, to study the QNMs of BHs that posses particular boundary conditions as being described in Sec. \ref{II}. To test the method, we have considered the
$4D$ pure \textsl{dS} and $5D$ \textsl{AdS}-Schwarzschild BHs.
Scalar field perturbations have been treated as oscillations in the frequency
domain of those static and symmetric backgrounds. In each geometry, the
perturbed scalar fields are reduced to $1D$ Schr\"{o}dinger like
wave equations with the associated effective potential. Imposing the required
boundary conditions given in Refs. \cite{IZSMAIN} and
\cite{Govindarajan:2000vq}, we have demonstrated how the FNNM derives the
QNMs: the resulting formula is Eq. \eqref{best}. After comparing our findings
with the previous results obtained by the analytical method  \cite{IZSMAIN}
and the superpotential approach (numerical) method \cite{Govindarajan:2000vq}
, it is seen that the all results are in good agreement with each other.
Therefore, FNNM is not only an alternative but an effective way for computing
the QNMs that are important for the stability of a BH and the late-time behavior of radiation from gravitationally
collapsing configurations. On the other hand, one may ask that how about for neural network solutions to the completely unknown problems. For such a case the architecture and training processes must be different than the FNNM that we employed here. In fact, such an architecture is more about interacting with ``\textit{experimental data}" like ``a\textit{ direct adaptive neural network method}`` \cite{example} in which the system took into account was described by an unknown NARMA model \cite{NARMA} and the FNNM was considered to learn the system. By taking the FNNM as a neural model, the control signals can easily be obtained by minimizing momentary difference or cumulative differences between a set point and the output of the FNNM. Since the training algorithm can ensure that the output of the FNNM approaches to the real system, then it can be demonstrated how the obtained control signals make the real system output as being close to the set point \cite {example2}. However, such an algorithm cannot be established without having experimental data on the BHs that we work with. Namely, with the development of technology related to the BHs, it might be possible to construct such a neural network algorithm.

Further work to determine the QNMs of rotating and/or higher/lower dimensional \textsl{dS/AdS}
spacetimes via the FNNM could therefore be interesting.
Besides, we aim to extend our analysis to the Dirac (e.g., the reader is referred to  this recent study \cite{fermion}) and Maxwell equations that
are formulated in the Newman-Penrose formalism \cite{mynp1,mynp2,mynp3} in the near future. Moreover, starting from Kerr BH, we also plan to analyze the QNMs \cite{myq1,myq2,myq3,myq4,myq5} of various stationary spacetimes.

\acknowledgments 
We are thankful to the Editor and anonymous Referees for their constructive suggestions and comments. 

\appendix
\textbf{\textcolor{black}{APPENDIX}}

Metric of the Reissner-Nordstr\"{o}m BH of mass $M$ and charge $Q$ is given by
\begin{equation}
d s^{2}=-\frac{\Delta}{r^{2}} d t^{2}+\frac{r^{2}}{\Delta} d r^{2}+r^{2}\left(d \theta^{2}+\sin ^{2} \theta d \varphi^{2}\right), \label{A1} \tag{A1}
\end{equation}
where $\Delta=r^{2}-2 M r+Q^{2}$. The locations of the event horizon and of the Cauchy horizon are $r_{+}=M+\sqrt{M^{2}-Q^{2}}$ and $r_{-}=M-\sqrt{M^{2}-Q^{2}}$, respectively. To investigate the bosonic perturbation of the Reissner-Nordstr\"{o}m BH, one should consider the scalar field $\Phi$, which obeys the Klein-Gordon equation (\ref{new2}), propagating in a Reissner-Nordstr\"{o}m BH geometry. For the chargeless case with an Ansatz $\Phi=R_{0}(r) Y_{j m}^{0}(\theta, \phi) e^{-i \omega t}$, the radial components of the fields can be found as follows \cite{izs1}:
\begin{equation}
\frac{d}{d r}\left(\Delta\frac{d R_{0}}{d r}\right)+\left(\frac{K^{2}}{\Delta}-\lambda\right) R_{s}=0, \label{A2} \tag{A2}
\end{equation}
where $K=\omega r^{2}$ and $\lambda=\ell(\ell+1)$ is a separation constant. If one makes the following transformation $f_{0}=r R_{0}$ and adopt the tortoise coordinate $r_{*}$ (defined here as $d r_{*} / d r=$ $\left.r^{2} / \Delta\right)$, the radial equation (\ref{A2}) recasts in
\begin{equation}
\frac{d^{2} f_{0}}{d r_{*}^{2}}+W_{0}\left(\omega, r_{*}\right) f_{0}=0, \label{A3} \tag{A3}
\end{equation}
where the complex function $W_{0}$ is given by
\begin{equation}
W_{0}\left(\omega, r_{*}\right)=\frac{\Delta}{r^{4}}\left[\frac{K^{2}}{\Delta}-2 \frac{M}{r}+2 \frac{Q^{2}}{r^{2}}-\lambda\right]. \label{A4} \tag{A4}
\end{equation}

By comparing Eqs. (\ref{new13}) and (\ref{A3}), one can easily derive the effective potential of the Reissner-Nordstr\"{o}m BH:
\begin{equation}
V^{\textsl{RN}}_{0}(r)=\frac{\Delta}{r^{4}}\left[
2 \frac{M}{r}-2 \frac{Q^{2}}{r^{2}}+\lambda\right]. \label{A5} \tag{A5}
\end{equation}

Following the Leaver's \cite{izs2,izs3} original continued fraction method (CFM), which was later improved by
Nollert \cite{izs4}, with the effective potential (\ref{A5}), Richartz and Giugno \cite{izs5} obtained the numerical values of the QNMs of the Reissner-Nordstr\"{o}m BH. Comparing the numerical results of Ref. \cite{izs5} with the results to be obtained from FNNM might be more meaningful and beneficial for the reader. For this purpose, we have created Table III, which obviously shows how the two methods produce very close values.

\begin{table}[h!]
\begin{ruledtabular}
\begin{tabular}{|p{1.8cm}|p{2cm}|p{2cm}|p{2cm}|p{2cm}|p{2cm}|}
$Q/M$ &0.01&  0.1 & 0.5 & 0.99 & 0.9999 \\
\hline
FNNM  & $0.110645-0.103976i$ & $0.111758-0.104364i$& $0.115843-0.10628i$ &$0.133596-0.094459i$ & $0.133507-0.094478i$  \\
\hline 
CFM &  $0.110457- 0.104896i $ & $ 0.110649  - 0.104938i$ & $ 0.115764  - 0.105751i $ & $ 0.133570 - 0.095641i $ & $0.133459- 0.095844i$ \\
\end{tabular}
\caption{QNM frequencies of the Reissner-Nordstr\"{o}m BH for the fundamental modes ($n=0$) obtained from the continued fraction method (CFM \cite{izs5}) and FNNM: Qualitative comparison of the two methods. For the sake of simplicity, in this illustrative example, we have considered the s-modes ($\ell$) of the chargeless ($q=0$) scalar fields. See Table I of Ref. \cite{izs5}.}%
\end{ruledtabular} \label{Tab3}
\end{table}

\end{document}